\documentclass[twocolumn,preprint]{aastex63}

\shorttitle{Encounter rate}
\shortauthors{Rawiraswattana \& Goodwin}
\graphicspath{{./}{figures/}}

\usepackage{graphicx}
\usepackage{amsmath}
\usepackage{amssymb}

\usepackage{bm}
\usepackage{multirow}
\usepackage{booktabs}

\newcommand{\encounter}{\rm e}
\newcommand{\stars}{\rm s}
\newcommand{\crossingtime}{\rm cr}
\newcommand{\mode}{\rm m}
\newcommand{\halfmass}{\rm h}
\newcommand{\mean}{\rm m}
\newcommand{\simulated}{\rm sim}
\newcommand{\estimated}{\rm est}

\newcommand{\dee}[1]{\mathrm{d}{#1}}
\newcommand{\rbk}[1]{\left(#1\right)}     

\begin{document}

\title{On encounter rates in star clusters}

\correspondingauthor{Simon P. Goodwin}
\email{s.goodwin@sheffield.ac.uk}

\author{Krisada Rawiraswattana}
\affiliation{Division of Physical Science, Faculty of Science,\\
Prince of Songkla University, 15 Kanjanavanich Road,\\
Hatyai, Songkhla 90110, Thailand}

\author[0000-0001-6396-581X]{Simon P. Goodwin}
\affiliation{Department of Physics and Astronomy,\\
University of Sheffield, Sheffield S3 7RH, UK}

\begin{abstract}

Close encounters between stars in star forming regions are important as they can perturb or destroy protoplanetary discs, young planetary systems, and stellar multiple systems.  We simulate simple, viralised, equal-mass $N$-body star clusters and find that both the rate and total number of encounters between stars varies by factors of several in statistically identical clusters due to the stochastic/chaotic details of orbits and stellar dynamics.  Encounters tend to rapidly `saturate' in the core of a cluster, with stars there each having many encounters, while more distant stars have none.  However, we find that the fraction of stars that have had at least one encounter within a particular distance grows in the same way (scaling with crossing time and half-mass radius) in all clusters, and we present a new (empirical) way of estimating the fraction of stars that have had at least one encounter at a particular distance.

\end{abstract}


\section{Introduction}\label{SEC:INTRODUCTION}

Young stars are commonly found with circumstellar discs \citep[e.g.][]{Hillenbrand:etal:1998,Lada:etal:2000,Haisch:etal:2000,Haisch:etal:2001}, and
these discs are thought to be where planet formation occurs.
Since most stars are formed in relatively dense environments \citep[e.g.][]{Lada:Lada:2003}, it is possible for the discs, and the on going planet formation process within, to be affected by close encounters between stars.

Simulations have shown that the effect of tidal perturbation from a stellar fly-by can range from slightly changing the density distribution in the disc to truncating or even destroying it \citep[e.g.][]{Clarke:Pringle:1993,Cuelloetal22}, depending on how close the encounter is.
This dynamical truncation, as well as photoevaporation \citep[e.g.][]{Concha-Ramirez:etal:2022}, and face-on accretion \citep{Wijnen:etal:2017}, can significantly affect the population of young stars with discs in the early stages of star formation.
Perturbations can also trigger disc instabilities \citep[e.g.][]{Thies:etal:2005,Thies:etal:2010} and may determine the population of planets forming in the disc \citep[][]{Ndugu:etal:2022}.
Another interesting effect of encounters on the disc is the misalignment between the rotational planes of the disc and the host star due to a non-coplanar encounter \citep[e.g.][]{Heller:1993,Larwood:1997}.
Encounters may alter already formed planetary systems, changing orbits  \citep[e.g.][]{Breslau+19}, or disrupting them \citep[e.g.][]{Parker+12}.  And encounters can similarly alter or destroy multiple stellar systems \citep[e.g.][]{Goodwin10,Reipurth14}.

Young stars with masses $\lesssim 1$ M$_{\sun}$ typically have discs with radii of a few hundreds of au \citep[e.g.][]{Andrews:Williams:2007}.
For the discs of those stars to be significantly perturbed in encounters, the periastron distance between the encountering stars needs to be less than $\sim 1000$ au.
Therefore, to understand how important encounters are in affecting discs/planet formation, a key question is how many young stars have encounters within $1000$ au?

There are two approaches one might take to finding the rates and numbers of encounters in some star cluster of interest.  The first is to perform $N$-body simulations of a variety of similar systems which is time consuming and computationally expensive \citep[e.g.][]{Parker+12,Craig+13}, the second would be to have some (ideally analytic) estimate to quickly get at least a `feel' for the expected values.

In this paper, we examine the encounter rate in a number of $N$-body simulations of bound star clusters.  We show that encounter {\rm rates} can vary by up to an order of magnitude between statistically identical clusters, but the fraction of stars that have had an encounter remains statistically the same.  We present an empirical way of estimating the fraction of stars in a cluster that have had at least one encounter within a particular distance.

\section{The encounter rate}\label{SEC:ENCOUNTER-RATE}

The number of encounters per unit time ($\varepsilon$) for a star seems like it should depend on several factors: the encounter distance of interest, some average number density of stars, and the typical velocity of stars.  The velocity of stars will affect the encounter rate by both changing how often stars encounter other stars, and also changing how effective gravitational focusing is.

\subsection{The standard method}\label{SEC:STANDARD-METHOD}

The most common method of calculating encounter times is based on the fundamental assumptions that a star is travelling through an effectively infinite, uniform density medium at a constant speed \citep[see e.g. the derivation in][note that these assumptions are perfectly adequate if one is interested in e.g. the Galactic disc]{Binney:Tremaine:2008}.

The encounter rate for any individual star is typically given by
\begin{equation}\label{EQN:STAR-ENCOUNTER-RATE}
    \varepsilon = 4\sqrt{\pi}n\sigma\rbk{r_{\encounter}^{2}+\frac{Gm}{\sigma^{2}}r_{\encounter}},
\end{equation}
where $n$ is the number density of the stars, $\sigma$ is the velocity dispersion, $r_{\encounter}$ is the closest distance during the encounter, $G$ is the gravitational constant, and $m$ is the typical mass of the stars \citep{Binney:Tremaine:2008}.  The second term in the brackets is associated with the gravitational focusing effect which deflects the trajectories and decreases the distance of closest approach for slow encounters or encounters between more massive stars.

For an ensemble of $N$ stars (e.g. a cluster), it seems reasonable to assume that the total encounter rate (total number of encounters per unit time, $\mathcal{E}$) scales with the total number of stars. Since each encounter involves two stars, the encounter rate should scale with $N/2$, i.e. $\mathcal{E} \simeq N\varepsilon/2$.

In convenient units where $r_{\encounter}$ is in au, $\sigma$ in km\,s$^{-1}$, $n$ in pc$^{-3}$, and $m$ in M$_{\sun}$, the encounter rate in a cluster of $N$ stars is then
\begin{equation}\label{EQN:CLUSTER-ENCOUNTER-RATE}
    \mathcal{E} = 8.5\times10^{-11}Nn\sigma\rbk{r_{\encounter}^{2}+886\frac{m}{\sigma^{2}}r_{\encounter}}\ \textrm{Myr}^{-1}.
\end{equation}

\bigskip

Therefore, it would seem that to calculate the rate of encounters, $\mathcal{E}$, at a particular distance of interest, $r_{\encounter}$, in an ensemble of $N$ stars, the correct values of (a) number density, $n$, and (b) velocity dispersion, $\sigma$, are required.  If there is a distribution of stellar masses, an appropriate value for $m$ must be taken.

While various assumptions that go into this simple calculation are clearly wrong for star clusters (e.g. moving through an effectively infinite uniform density medium at a constant speed) one might think that some simple variation on this approach might work (e.g. taking some appropriate average speed and density).  However, we will show that this approach in star clusters gives an often wrong, and an always misleading, `answer'.

\subsection{What do we want to know?}

It is important to clarify what we want to know about encounters in a cluster.
In most cases, what we would like is an estimate of {\em what fraction of stars have had a close encounter} as this tells us the relative levels of disc/planetary system/multiple system  perturbation/destruction.  It is important to remember that this is {\em not} what an estimate of an encounter rate gives without a further assumption of how the encounters are distributed between stars.

As an example let us take a cluster that we shall examine in detail later: an $N = 300$, $M = 300$ M$_{\sun}$, equal-mass (so $m = 1$ M$_{\sun}$) virialised Plummer sphere cluster with half-mass radius $r_{\halfmass} = 0.5$ pc.  If we want to know the number of stars that have had an encounter within, e.g. $r_{\encounter} = 1000$ au, we can calculate that $\mathcal{E} \sim 25$ Myr$^{-1}$ by taking the values for the velocity dispersion and half-mass density of this cluster and putting them into equation \eqref{EQN:CLUSTER-ENCOUNTER-RATE}.

If we assume this encounter rate estimate is correct, to calculate how many stars have had an encounter after some time we need to make a further assumption that encounters are random so that after $2$ Myr there will be 50 stars that have had an encounter, and after $10$ Myr $250$ stars (ie. $> 80$ per cent) will have had an encounter.  (One could be somewhat more sophisticated and estimate as the encounter fraction starts to approach unity how many stars have had zero, one, two etc. encounters.)

In assuming encounters are random, this calculation ignores that encounters are much more likely to occur in the core, and that after a few crossing times some stars in the core are likely to have had multiple encounters, while those in the halo may have had none.  Indeed, when stated like this, this approach does seem extremely naive and it would be surprising if it gave the correct answer.
 \bigskip

\section{Simulations}

We investigate encounter rates in clusters by performing $N$-body simulations in which we can record individual encounters, the stars involved, and their distances of closest approach.  The simulations we report here are of the simplest bound systems: virialised Plummer spheres of equal-mass stars.

We simulate $N = 300$ and $N = 600$ virialised Plummer spheres \citep{Plummer:1911} initialised by the method described in \citet{Aarseth:etal:1974} with initial half-mass radii of $0.5$, $0.75$ and $1$ pc.  Simulations are run only with equal-mass stars so that we can ignore any complicating effects of mass spectra.

The number of stars ($N$), the stellar mass ($m$), the initial half-mass radius ($r_{\halfmass}$), and the label of simulated clusters are shown in Table \ref{TAB:SIMULATION-LABEL}.  Clusters with $r_{\rm h} = 0.5$ pc are truncated at $3$ pc, those with $r_{\rm h} = 0.75$ pc are truncated at 5 pc, and $r_{\rm h} = 1$ pc clusters are truncated at 7 pc
so that they have approximately the same relative sizes.

\begin{table}
    \renewcommand{\arraystretch}{1.2}
    \centering
    \caption{The properties of cluster ensembles.  Each ensemble contains 10 clusters with different random number seeds (labelled {\tt A} to {\tt J}) with the same number of stars ($N$), stellar masses ($m$), half-mass radii ($r_{\rm h}$), and crossing time ($t_{\rm cr}$).  The ID of a simulation contains information on the initial conditions: \texttt{N3} or \texttt{N6} are $N = 300$ and $N = 600$ respectively, \texttt{SM} stands for single-mass, and after \texttt{R} is the initial half mass radius of the cluster.}
    \label{TAB:SIMULATION-LABEL}
    \begin{tabular}{@{ }lcccc@{ }}
        \hline\hline
        Cluster IDs & $N$ & $m$ (M$_{\sun}$)  & $r_{\halfmass}$ (pc) & $t_{\rm cr}$ (Myr) \\
        \midrule
        \texttt{N3SMR050-A} to \texttt{J}   & \multirow{3}{*}{$300$}    & \multirow{3}{*}{$1$}      & $0.5$ & 1.0 \\
        \texttt{N3SMR075-A} to \texttt{J}   &                           &                           & $0.75$ & 1.9 \\
        \texttt{N3SMR100-A} to \texttt{J}   &                           &                           & $1$ & 3.1\\
        \midrule
        \texttt{N6SMR050-A} to \texttt{J}   & \multirow{1}{*}{$600$}    & \multirow{1}{*}{$1$}      & $0.5$ & 0.74\\
        \bottomrule
    \end{tabular}
\end{table}

All simulations are run for $10$ Myr using our own $N$-body code.
The code uses a forth-order Hermite scheme \citep{Makino:Aarseth:1992} as the integrator. We keep the energy error well below $10^{-4}$ by employing an adaptive timestep, i.e. using equation (7) from \citet{Makino:Aarseth:1992} with parameter $\eta = 4 \times 10^{-4}$ for $N = 300$ clusters and $\eta = 1 \times 10^{-4}$ for $N = 600$ clusters.
We also use block timestepping for $N = 600$ runs to speed up the calculations.

The separation between any pair of stars is monitored at every timestep. Once two stars are closer to each other than $1000$ au, whether they are bound or unbound, they are considered as having close encounter.
During this period, the closest separation is recorded and taken as the encounter distance once the stars move away from each other beyond $1000$ au. In binaries multiple `encounters' will occur, if the separation stays below $1000$ au this is only counted as one `encounter'. This is to prevent hard binaries inflating the close encounter rate, however as we discuss below hard binaries form extremely rarely in our simulations.


Simulations are run with a gravitational softening length of $0.01$ au to avoid collisions or computationally expensive very close encounters.  This is only of importance for the details of extremely close encounters at $\ll 10$ au which is much closer than the vast majority of encounters and at a distance that would completely disrupt discs or  planetary systems.

\subsection{Number density and velocity dispersion}

It would seem reasonable to assume that encounter rates and fractions should depend in some way on both number density and velocity dispersion.  Below we go into some detail on how the number density and velocity dispersion of a cluster might be quantified.  This is important to show that the encounter rates we measure in simulations disagree with simple calculations because the assumptions that underlie them are wrong for clusters, rather than that we are using the `wrong' values for number density or velocity dispersion, or not accounting for how they change with time.  A reader happy to take our word for this can skip the details below.

\subsubsection{Number density}

The derivation of equation \eqref{EQN:CLUSTER-ENCOUNTER-RATE} assumes that stars are uniformly distributed and so $n$ is constant in time and space.  This is a reasonable assumption for e.g. encounters in the Galactic disc, but not for encounters in a cluster (or any region where the number density varies on short length-scales).

There are a number of ways one could quantify some average number density, and they can result in very different values for the estimated encounter rate.

The first average number density we use is simply calculated from the half-mass radius of the cluster ($r_{\halfmass}$):
\begin{equation}\label{EQN:HALF-MASS-NUMBER-DENSITY}
    n_{\halfmass} = \frac{3N}{4\pi{}r_{\halfmass}^{3}}.
\end{equation}

The second average number density is the mean number density defined by
\begin{equation}\label{EQN:MEAN-NUMBER-DENSITY-1}
    n_{\mean} = \frac{\int_{0}^{\infty}nf\dee{r}}{\int_{0}^{\infty}f\dee{r}} = \int_{0}^{\infty}nf\dee{r},
\end{equation}
where $n$ and $f$ are the number density and the probability density function of the distance of stars from the centre of mass of the cluster ($r$).
In practice, we can approximate the integral by
\begin{equation}\label{EQN:MEAN-NUMBER-DENSITY-2}
    n_{\mean} \simeq \Delta{}r\sum_{i = 1}^{N_{\rm bin}}n_{i}f_{i},
\end{equation}
where $\Delta{}r$ and $N_{\rm bin}$ are the size and the number of bins of  stellar distances from the centre of mass of the cluster.

Theoretically, the probability density function is defined as $f = \dee{P}/\dee{r}$, where $\dee{P}$ is the probability of finding stars at distance between $r$ and $r+\dee{r}$ from the centre of mass of the cluster.
But practically, the value of $f_{i}$ in equation \eqref{EQN:MEAN-NUMBER-DENSITY-2} may be obtained from
\begin{equation}\label{EQN:DISTANCE-PDF}
    f_{i} = \rbk{\frac{\Delta{}P}{\Delta{}r}}_{i} = \frac{N_{i}}{N\Delta{}r},
\end{equation}
where $N_{i}$ is the number of stars in the $i^{\rm th}$-bin and $N$ is the number of stars in the cluster.
The number density $n_{i}$ in equation \eqref{EQN:MEAN-NUMBER-DENSITY-2} is related to $f_{i}$ via
\begin{equation}\label{EQN:ITH-NUMBER-DENSITY}
    n_{i} = \rbk{\frac{\Delta{}N}{\Delta{}V}}_{i} = \frac{N}{4\pi{}r_{i}^{2}}\rbk{\frac{\Delta{}P}{\Delta{}r}}_{i} = \frac{Nf_{i}}{4\pi{}r_{i}^{2}},
\end{equation}
where $\Delta{}V$ is a spherical volume element containing $\Delta{}N$ stars.
Substituting equations \eqref{EQN:DISTANCE-PDF} and \eqref{EQN:ITH-NUMBER-DENSITY} in \eqref{EQN:MEAN-NUMBER-DENSITY-2} gives
\begin{equation}\label{EQN:MEAN-NUMBER-DENSITY-3}
    n_{\mean} \simeq \frac{1}{4\pi{}N\Delta{}r}\sum_{i = 1}^{N_{\rm bin}}\frac{N_{i}^{2}}{r_{i}^{2}}.
\end{equation}

\bigskip

The half-mass number density is often used as it is simple to calculate, the more complex mean number density has the advantage of including information on the full density distribution of the cluster.
We also note here that the half-mass radius is often used as a characteristic radius as it remains roughly constant over the long-term evolution of a cluster \citep{Aarseth:etal:1974}, however the half-mass radius does fluctuate, especially at early times and so even the half-mass density changes (sometimes by factors of several).

\subsubsection{Velocity dispersion}

The velocity dispersion ($\sigma$) in equation \eqref{EQN:STAR-ENCOUNTER-RATE} and \eqref{EQN:CLUSTER-ENCOUNTER-RATE} comes from the assumption that the velocity distribution of the stars in the cluster is Maxwellian.
From the Maxwell-Boltzmann distribution, the velocity dispersion is related to the mode of the distribution ($v_{\mode}$) by $\sigma = v_{\mode}/\sqrt{2}$.
The mode can simply be determined by constructing the velocity histogram and then fitting it with a polynomial regression to find the position of the peak.

It should be noted that this way of finding the velocity dispersion is only possible when all (3D) velocities are well known.  In any observation the value of $\sigma$ is either `guessed' by assuming virial equilibrium, or observed in either 1D (radial velocities) or 2D (proper motions) with usually quite significant errors and biases (e.g. binary inflation, see \citet{Cottaar:etal:2012}).

\subsubsection{Time-averaging}

There are two ways to calculate the number density and velocity dispersion to use in equation \eqref{EQN:CLUSTER-ENCOUNTER-RATE}.  One is to take the values of $n$ or $\sigma$ calculated instantaneously at the end of the simulation, the other is to take a time average.

In a simulation it is possible to calculate full 3D time-averaged values for various quantities.  However, to estimate encounter rates in an observed region or for a single snapshot only the current values for any quantity can be calculated, and even they might be uncertain/guesstimated (e.g. no velocity data is available and only 2D positions).

For later reference, Table 2 shows the initial, time-averaged and final (i.e. those at 10 Myr) values of $\sigma$, $n_{\halfmass}$, and $n_{\mean}$ for all simulations.

\section{Results}

In our simulations we follow all encounters at distances of $<1000$ au.
We record when the encounter occurred, which two stars were involved, and the distance of closest approach.
This allows us to find the encounter rate within a particular distance (i.e. what equation \eqref{EQN:CLUSTER-ENCOUNTER-RATE} attempts to estimate), and the number of stars that have had such an encounter -- something equation \eqref{EQN:CLUSTER-ENCOUNTER-RATE} cannot tell us without further assumptions, but is often what we wish to know.

\subsection{Comparing an \texorpdfstring{$N = 300$}{} and an \texorpdfstring{$N = 600$}{} cluster}

We start by comparing encounter rates at various distances in two clusters with $N = 300$ and $N = 600$ equal-mass stars.  Both are initially virialised Plummer Spheres with $r_{\halfmass} = 0.5$ pc, with stars each of mass $1$ M$_{\sun}$.

\begin{figure}[ht!]
    \includegraphics[angle=270,width=\columnwidth]{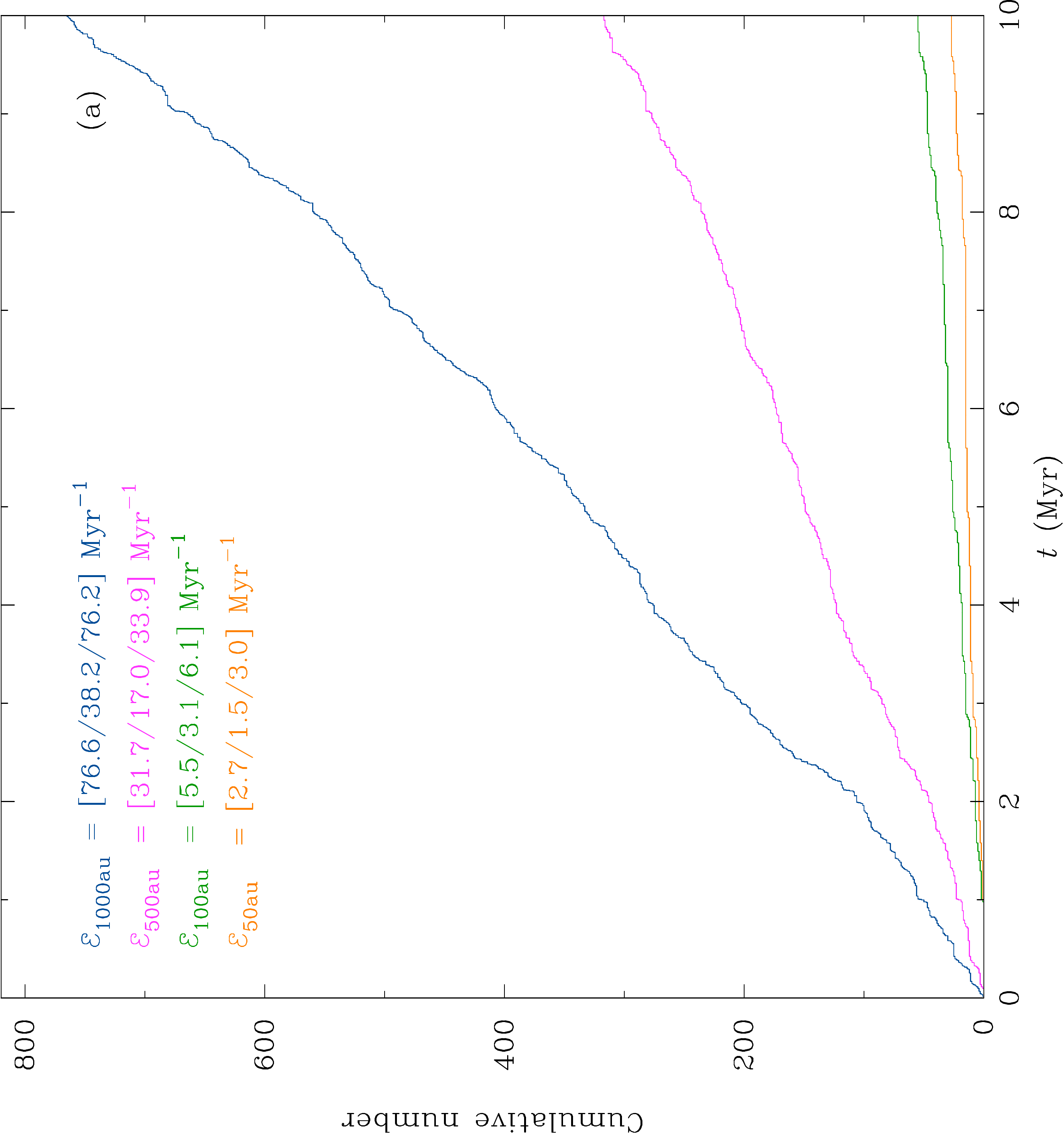}
    \includegraphics[angle=270,width=\columnwidth]{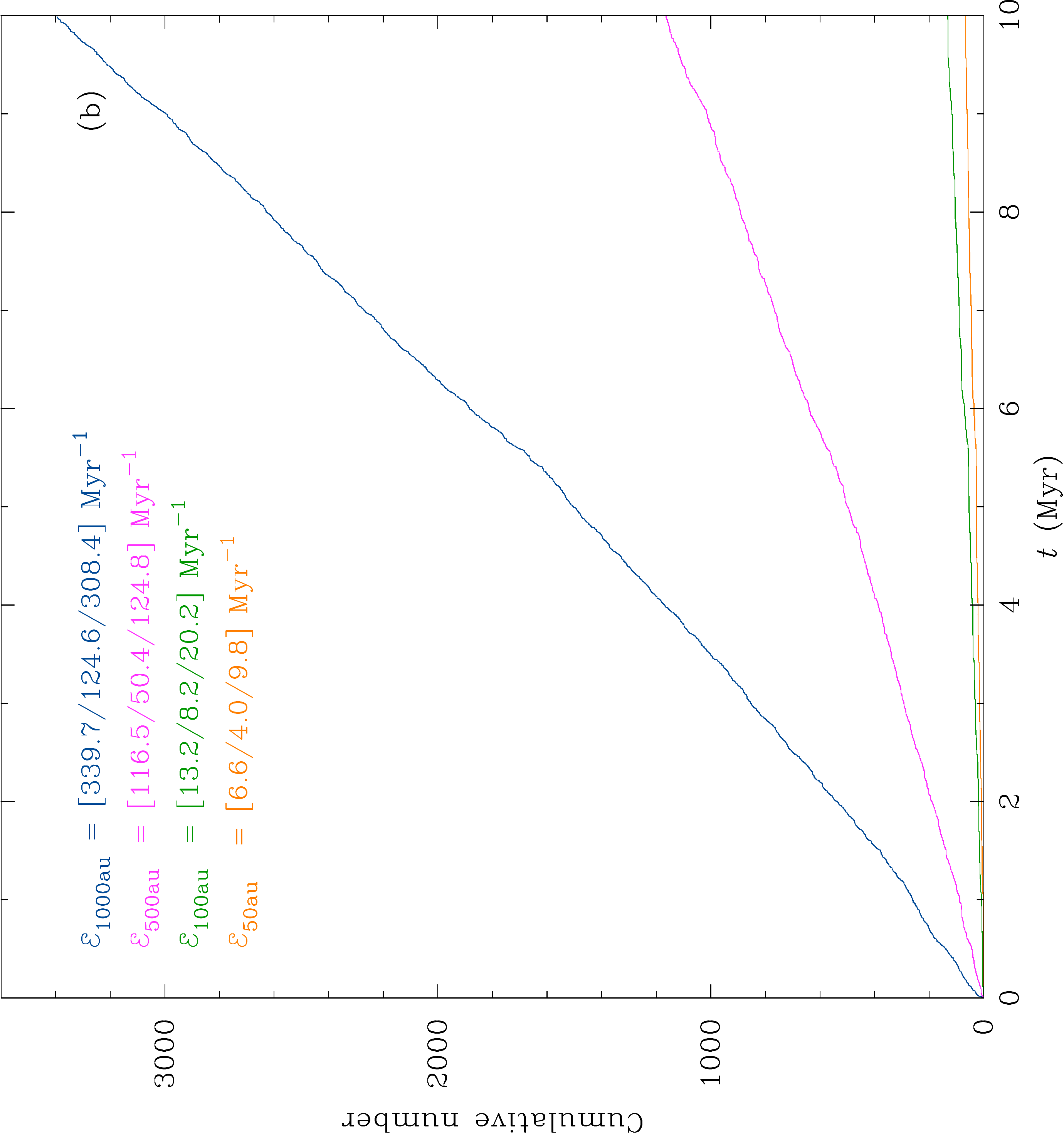}
    \caption{
        Cumulative numbers of encounters in runs  \texttt{N3SMR050-A} (top panel, a) and \texttt{N6SMR050-A} (bottom panel, b).
        The four curves in each panels are for encounter distances $r_{\encounter} < 50$ (orange), $100$ (green), $500$ (magenta), and $1000$ (blue) au, from bottom to top.
        Numbers in the square brackets in the top left are the encounter rates $[\mathcal{E}_{\simulated}/\mathcal{E}_{\estimated}(n_{\halfmass})/\mathcal{E}_{\estimated}(n_{\mean})]$ (see text).
    }
    \label{FIG:CUMULATIVE-NUMBER-SM}
\end{figure}

Figure \ref{FIG:CUMULATIVE-NUMBER-SM} shows the cumulative number of encounters over $10$ Myr in the $N = 300$ (top panel) and $N = 600$ (bottom panel) clusters.
In each panel the lines from bottom-to-top are the cumulative numbers of encounters at $r_{\encounter} < 50$ (orange), $100$ (green), $500$ (magenta), and $1000$ (blue) au respectively.
At the top left of each sub-figure are three numbers for each of the encounter distances: the first is the actual encounter rate (Myr$^{-1}$) as found in the simulation, the next two are time-averaged estimates that we will return to later, but note for now that all three numbers are often quite different.  The two simulations shown are \texttt{N3SMR050-A} (top) and \texttt{N6SMR050-A} (bottom).

Figure \ref{FIG:CUMULATIVE-NUMBER-SM} shows a number of features one might expect.
\begin{enumerate}
    \item The total number of encounters grows roughly linearly with time (in these two clusters at least).
    \item The number of encounters at different distances scales very roughly with $r_{\encounter}^{2}$ (e.g. for $N=300$ after $10$ Myr there have been $317$ encounters at $r_{\encounter} < 500$ au, and $27$ at $< 50$ au).
    \item Increasing both $N$ (and therefore also $n$) by a factor of two results in about $4$ times more encounters (e.g. $317$ when $N = 300$, and $1165$ when $N = 600$ at $r_{\encounter} < 500$ au).
\end{enumerate}

The second and third numbers in the top left are the estimates of encounter rate as calculated from equation \eqref{EQN:CLUSTER-ENCOUNTER-RATE} using the time-averaged values of the half-mass number density ($n_{\halfmass}$) and the mean number density ($n_{\mean}$) respectively.

In both cases using $n_{\halfmass}$ under-estimates the number of encounters by a factor of $\sim 2$.  Using $n_{\mean}$ seems better, often giving a reasonable estimate (but sometimes being off by a factor of $\sim 2$).  So at a first glance at just these two simulations one might consider that using $n_{\mean}$ often provides a reasonable estimate of the encounter rate in a cluster.

However, we have only compared two simulations which just happened to be those labelled  {\tt A} in our ensembles.  As we show below, when we look at the whole ensemble the picture becomes {\em much} more complicated, and this emphasises the importance of looking at ensembles of simulations when dealing with $N$-body systems.

\begin{table*}
    \renewcommand{\arraystretch}{1.2}
    \centering
    \scriptsize
   \caption{
        {
            The first three triple columns are the initial (Ini.), time averaged (Avg.) and final (End) values of the velocity dispersion ($\sigma$), the half-mass number density ($n_{\halfmass}$) and the mean number density ($n_{\mean}$) of clusters \texttt{N3SMR050-A..J} (initial conditions in Table \ref{TAB:SIMULATION-LABEL}).
        }
        In the last triple column are the analytic estimates of the encounter rates at $r_{\encounter} < 1000$ au, using the { average} half-mass number density ($\mathcal{E}_{\estimated}(n_{\halfmass})$), and the {average} mean number density ($\mathcal{E}_{\estimated}(n_{\mean})$), compared with the actual values of the encounter rate measured in the simulations ($\mathcal{E}_{\simulated}$).
        {
            Note that the encounter rate can also be calculated from the initial or final values of $\sigma$, $n_{\halfmass}$ and $n_{\mean}$.
        }
    }\label{TAB:NUMBER-DENSITY}
    \begin{tabular}{@{ }l@{ }c@{ }c@{ }c@{ }c@{ }c@{ }c@{ }c@{ }c@{ }c@{ }c@{ }c@{ }c@{ }}
        \hline\hline
        \multirow{2}{*}{Cluster}    & \multicolumn{3}{c}{$\sigma$ (km\,s$^{-1}$)}
                                    & \multicolumn{3}{c}{$n_{\halfmass}$ (pc$^{-3}$)}
                                    & \multicolumn{3}{c}{$n_{\mean}$ (pc$^{-3}$)}
                                    & \multicolumn{3}{c}{$\mathcal{E}$ at $r_{\encounter} < 1$ kau (Myr$^{-1}$)}\\
                                    \cmidrule(rr){2-4}\cmidrule(rr){5-7}\cmidrule(rr){8-10}\cmidrule(rr){11-13}
                                    & \multicolumn{1}{c}{Ini.} & \multicolumn{1}{c}{Avg.} & End
                                    & \multicolumn{1}{c}{Ini.} & \multicolumn{1}{c}{Avg.} & End
                                    & \multicolumn{1}{c}{Ini.} & \multicolumn{1}{c}{Avg.} & End
                                    & \multicolumn{1}{c}{$\mathcal{E}_{\estimated}(n_{\halfmass})$}
                                    & \multicolumn{1}{c}{$\mathcal{E}_{\estimated}(n_{\mean})$}
                                    & $\mathcal{E}_{\simulated}$ \\
        \midrule
        \texttt{N3SMR050-A} & $0.57$ & $0.50\pm0.04$ & $0.55$ & $541$ & $662\pm144$ & $786$ & $1925$ & $1320\pm475$ & $1650$ & $38\pm8$ & $76\pm28$ & $77$\\
        \texttt{N3SMR050-B} & $0.50$ & $0.51\pm0.05$ & $0.43$ & $543$ & $634\pm112$ & $693$ & $1735$ & $1018\pm482$ & $735$ & $36\pm7$ & $58\pm28$ & $58$\\
        \texttt{N3SMR050-C} & $0.55$ & $0.49\pm0.04$ & $0.43$ & $557$ & $382\pm66$ & $247$ & $1535$ & $950\pm780$ & $2402$ & $22\pm4$ & $56\pm46$ & $131$\\
        \texttt{N3SMR050-D} & $0.57$ & $0.54\pm0.04$ & $0.57$ & $571$ & $507\pm92$ & $474$ & $1877$ & $1396\pm948$ & $1313$ & $28\pm5$ & $78\pm53$ & $143$\\
        \texttt{N3SMR050-E} & $0.57$ & $0.53\pm0.06$ & $0.49$ & $577$ & $647\pm120$ & $475$ & $964$ & $1232\pm413$ & $1813$ & $36\pm7$ & $69\pm24$ & $60$\\
        \texttt{N3SMR050-F} & $0.53$ & $0.53\pm0.05$ & $0.42$ & $571$ & $682\pm102$ & $532$ & $766$ & $669\pm174$ & $484$ & $38\pm6$ & $38\pm10$ & $30$\\
        \texttt{N3SMR050-G} & $0.49$ & $0.50\pm0.05$ & $0.50$ & $577$ & $669\pm130$ & $480$ & $1906$ & $1291\pm537$ & $1158$ & $39\pm8$ & $75\pm31$ & $90$\\
        \texttt{N3SMR050-H} & $0.58$ & $0.51\pm0.04$ & $0.48$ & $555$ & $582\pm125$ & $551$ & $1435$ & $1710\pm723$ & $1100$ & $33\pm7$ & $98\pm42$ & $112$\\
        \texttt{N3SMR050-I} & $0.57$ & $0.50\pm0.04$ & $0.43$ & $549$ & $437\pm110$ & $228$ & $1029$ & $956\pm497$ & $531$ & $25\pm6$ & $55\pm29$ & $208$\\
        \texttt{N3SMR050-J} & $0.50$ & $0.50\pm0.05$ & $0.46$ & $577$ & $441\pm103$ & $375$ & $2359$ & $1583\pm917$ & $978$ & $26\pm6$ & $92\pm53$ & $198$\\
        \texttt{N3SMR050-A..J} & $0.54\pm0.03$ & $0.51\pm0.05$ & $0.48\pm0.05$ & $562\pm14$ & $564\pm110$ & $484\pm175$ & $1553\pm506$ & $1213\pm595$ & $1216\pm603$ & $32\pm7$ & $69\pm34$ & $111\pm60$\\
        \bottomrule
    \end{tabular}
\end{table*}

\subsection{An ensemble of statistically identical clusters}

We now consider all ten clusters in our $N = 300$ equal-mass stars, and a half-mass radius of $r_{\halfmass} = 0.5$ pc ensemble.  The only difference between these clusters is the random number seed used to generate the initial positions and velocities.  Therefore one would expect that the encounter rates in each would be similar  - and ideally be close to an analytic estimate.

\begin{figure}
    \includegraphics[angle=270,width=\columnwidth]{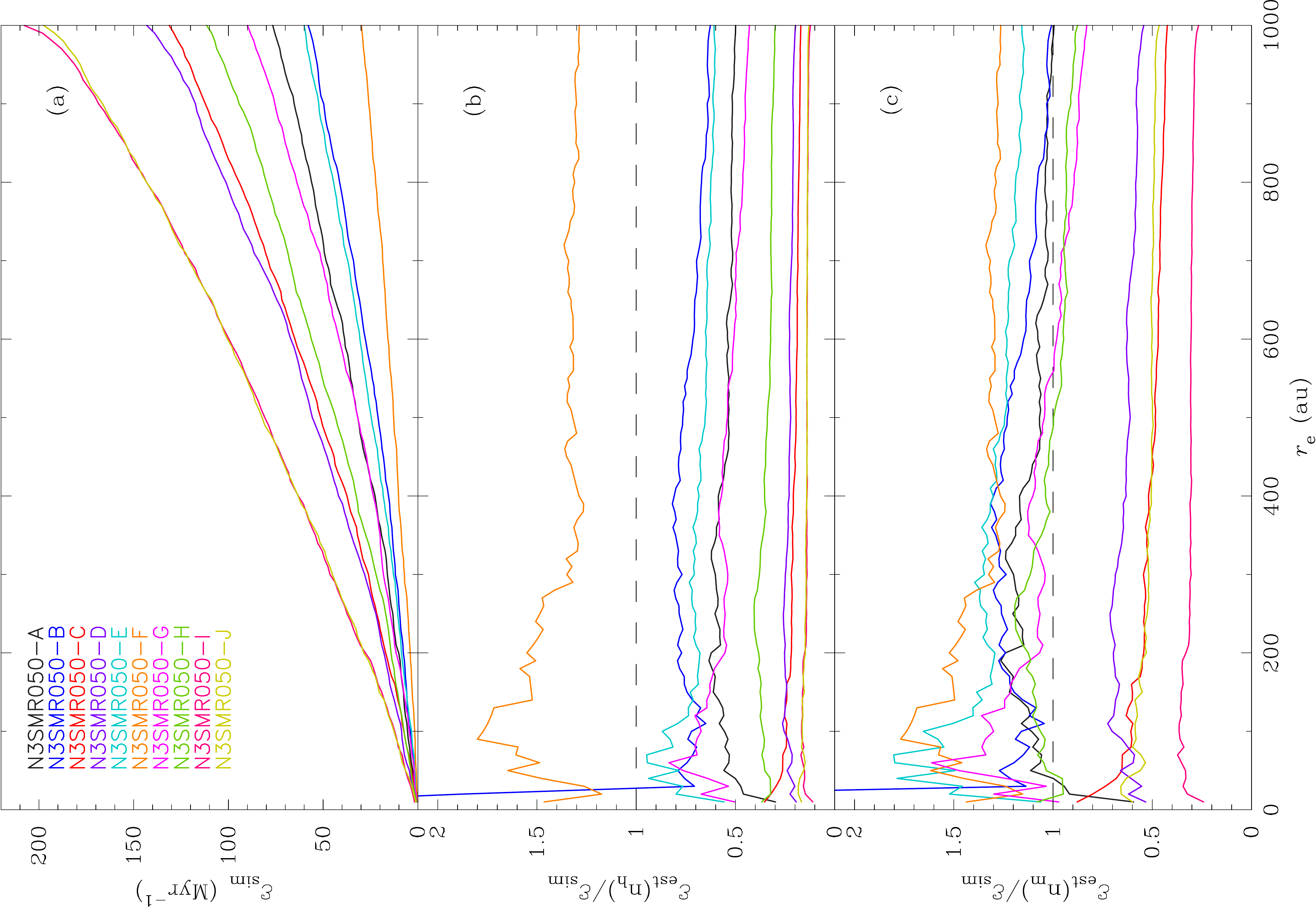}
    \caption{Top panel: the final encounter rates ($\mathcal{E}_{\simulated}$) against the encounter distance ($r_{\encounter}$)  from each cluster in the $N=300$, $r_{\rm h}=0.5$ pc ensemble.  Each cluster has a different colour as shown in the top left. Middle and bottom panels: the ratio of the analytic estimate to the actual encounter rate using the half-mass density (middle panel), and mean density (bottom panel).
    }
    \label{FIG:ENCOUNTER-RATES-N3SM}
\end{figure}

The top panel of Fig. \ref{FIG:ENCOUNTER-RATES-N3SM} shows the final encounter rates ($\mathcal{E}_{\simulated}$) for each of our identical clusters measured after $10$ Myr from $r_{\encounter} = 0$ to $1000$ au.\footnote{The exact values of encounter rates at separations of less than a few au may be affected by our softening, but a close encounter did happen, we just might not be able to trust the distance of closest approach too precisely.}  The simulation shown in the top panel of Fig. \ref{FIG:CUMULATIVE-NUMBER-SM} is \texttt{A} which is the black line towards the middle-bottom of all the lines.

The most important thing to note about the top panel of Fig. \ref{FIG:ENCOUNTER-RATES-N3SM} is the total encounter rates after 10 Myr varies very significantly between clusters with a difference of almost an order of magnitude.  There appears to be no `typical' clusters and some outliers -- just a seemingly random spread in encounter rates between clusters that are initially statistically identical.

Interestingly, all of the curves in the top panel of Fig. \ref{FIG:ENCOUNTER-RATES-N3SM} follow a distribution that goes roughly as $r_{\rm e}^2$ suggesting that the distribution of encounter distances is what would be expected for unbound encounters.
However, the distribution can slightly deviate from $r_{\rm e}^2$, often due to three-body encounters between a single star and a binary (which has formed during the simulation). These encounters can cause an increase in the encounter rate at $r_{\rm e} \sim 1000$ au, as can be most obviously seen in cluster \texttt{I} (dark red line at the top of the figure).

The other panels of Fig. \ref{FIG:ENCOUNTER-RATES-N3SM}
show the difference between the analytically estimated encounter rates and the actual encounter rates for each cluster.  In the middle panel the half-mass density ($n_{\rm h}$) is used for the estimate, and in the bottom panel it is the mean number density ($n_{\rm m}$).  A good match to the analytic estimate is the black dashed line at a ratio of unity.

Using the half-mass density never gives a good estimate, and can be wrong by an order of magnitude.  Using the mean density is {\em slightly} better - three or four clusters stay reasonably close to unity, but most clusters are always wrong by factors of several.

One obvious explanation would be that different clusters have changed significantly (some expanding and some contracting?).

In Table \ref{TAB:NUMBER-DENSITY} we give the initial, time averaged and final values of $\sigma$, $n_{\halfmass}$, and $n_{\mean}$ for all ten clusters, as well as the final cumulative encounter rate.  The averages and variance of each quantity over all the clusters are given in the bottom line.

The velocity dispersions ($\sigma$) are very similar between all clusters, but measures of density can change significantly over time with quite different time averaged and final values, and between different measures (half mass or mean).

However, these variations do not seem to correlate with the vastly different encounter rates.  The last three columns show the encounter rates estimated with $n_{\halfmass}$ and $n_{\mean}$ and then the actual encounter rates from the simulations.  Only once (cluster {\tt F}) does $n_{\halfmass}$ get close to predicting the actual encounter rate.  Estimates using $n_{\mean}$ are reasonable for 5/10 of the clusters, but very wrong for 5/10.  We think this was seen by \citet{Craig+13} who note that their simulations had far more encounters than one might expect, but substructure complicated their analysis.

What is particularly interesting is that there is no systematic change in encounters rates with any measure of density or how they evolve.  Cluster {\tt F} has the lowest final mean density (484 pc$^{-1}$) and the lowest encounter rate (30 Myr$^{-1}$), but cluster {\tt I} has the second lowest final mean density (531 pc$^{-1}$), but the highest encounter rate (208 Myr$^{-1}$).  Clusters {\tt B} and {\tt E} have almost the same encounter rate, but final mean densities that are different by a factor of over two (735 and 1813 pc$^{-1}$).

\begin{figure}
    \includegraphics[angle=270,width=\columnwidth]{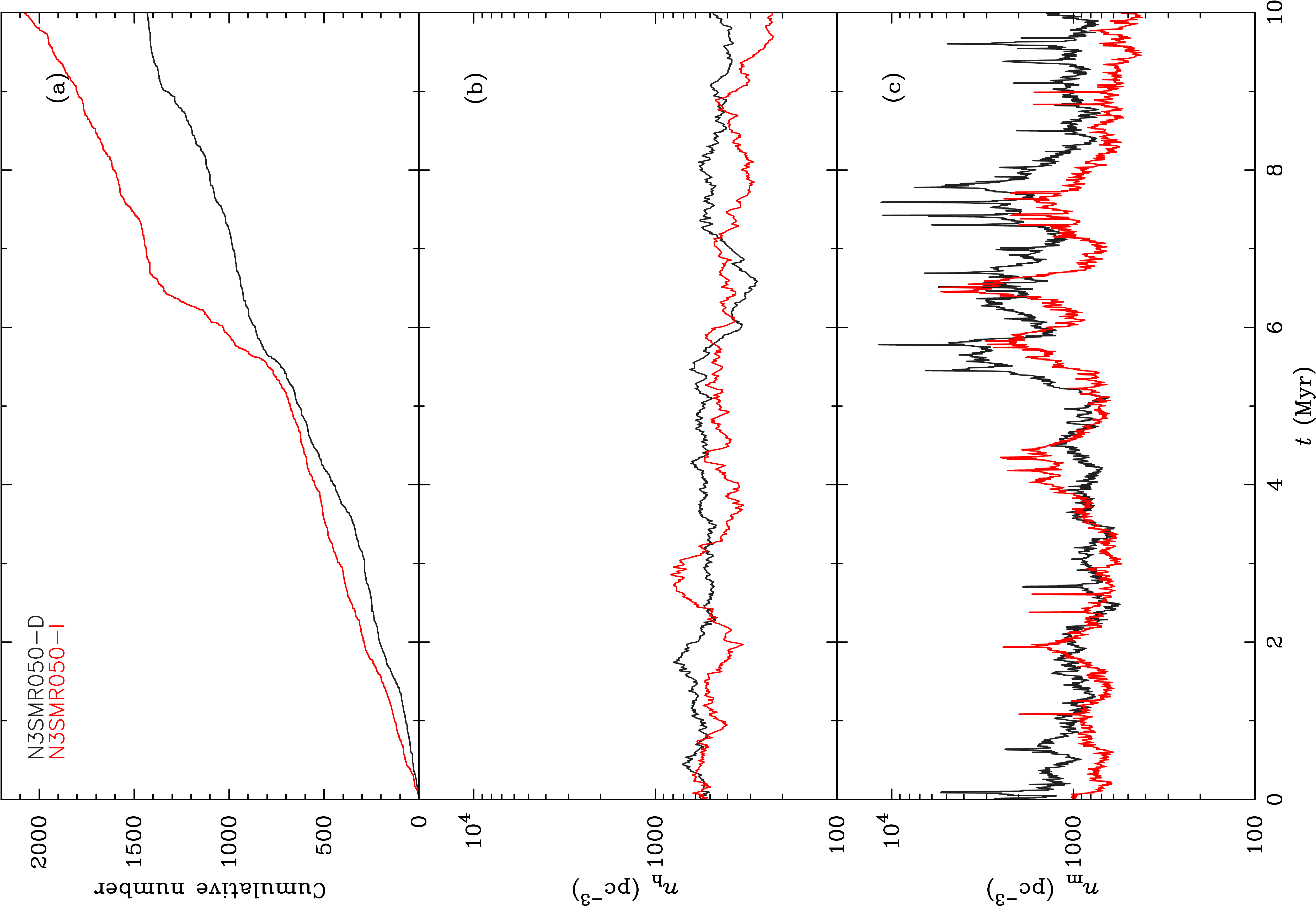}
    \caption{The evolution of encounter rates and density for cluster {\tt N3SMR050-D} (black lines) and cluster {\tt N3SMR050-I} (red lines).  Top panel (a): the number of encounters with time for $r_{\rm e} < 1000$ au. Middle panel (b): the half-mass density
    $n_{\halfmass}$.  Bottom panel (c): the mean density $n_{\mean}$.
    }
    \label{FIG:DENSITY-EVOLUTION-N3SM}
\end{figure}

One might think that maybe there was some extreme deviation in density at some point in time that the time averaged densities do not properly include, however this is not the case.
In Fig. \ref{FIG:DENSITY-EVOLUTION-N3SM} we show the numbers of encounters with $r_{\rm e} < 1000$ au (top panel), half-mass density (middle panel), and mean density (lower panel) for clusters {\tt D} and {\tt I} (cf. Fig. \ref{FIG:CUMULATIVE-NUMBER-SM}).

Cluster {\tt D} (black line) shows a relatively linear increase in encounter numbers to end with $\sim 1400$ encounters within 10 Myr.  Cluster {\tt I} (red line) is similar to cluster {\tt D} until a period between 6--7 Myr when the encounter rate increases significantly.

There is no reason to think that the increased encounter rate in cluster {\tt I} is due to density variations however.  The middle panel shows that both cluster's half-mass densities are very similar, and both fairly constant and declining slightly.  The bottom panel shows more variation in the mean densities with short-lived fluctuations of factors of a few, but both clusters show this behaviour, and, if anything,  cluster {\tt D} has higher densities.  There are fluctuations in the mean density of cluster {\tt I} around when the encounter rate increases significantly,  but there are others when it does not.

Examination of the data shows that the large numbers of encounters in cluster {\tt I} at 6--7 Myr is due to a few pairs of stars having multiple self-encounters in weakly bound pairs \citep[cf.][]{Moeckel:Clarke:2011}.

In all the ensembles of statistically identical clusters we have run we find no systematic relationship between any measure of cluster density and encounter rates (apart from occasionally in just a few of the clusters, but these could be chance given that many fluctuations do not correlate).

\subsubsection{Binaries}

All our simulations start with no binaries.  An interesting question is how many binaries can form, and how they might alter the evolution.

Soft binaries are extremely easy to form \citep[see][]{Moeckel:Clarke:2011}, and can inflate the encounter rate.  Any wide binary with periastron below 1000au and apastron above 1000au will be included as multiple encounters - however such binaries are extremely soft and short lived in our simulations (this was seen in cluster {\tt I}).

Hard binaries, however, are {\em much} more difficult to form.  The soft binaries we find are very weakly bound and can appear due to fluctuations in the global potential.  However, to form a hard, long-lived, binary system requires a three-body encounter as the third body is needed to carry-away the excess energy \citep[see][]{Goodman:Hut1993}.  A back-of-the-envelope calculation suggests hard binary formation should be rare in our clusters, and an examination of the simulations finds only a few  hard binaries have managed to form (one every few simulations, and never more than one in a simulation).

\subsection{The number of stars having had an encounter}

We clearly see that the {\em total number of encounters} between stars at any particular distance can be different by maybe an order of magnitude in initially statistically identical clusters, in a way that cannot be explained by density fluctuations.

The encounter rates we have shown in Fig. \ref{FIG:ENCOUNTER-RATES-N3SM} and Table \ref{TAB:NUMBER-DENSITY} are the number of times two stars come closer together than a particular distance.  However, this measure does not include information on if a particular star, or a particular pair of stars, have had multiple encounters.

Star clusters have a density distribution with a high density core and increasingly lower density as one moves outwards, and a Plummer profile is a reasonable approximation to young, bound clusters.

Within this density distribution stars can have a variety of orbits (which can change after encounters).  Some stars will spend a significant amount of their time in the high density core, some will spend most of their time in the low density halo, and various combinations in between (orbits can be radial or circular etc. and can change over time).

This means that each individual star will have a unique encounter history.  Those that spend a lot of time in the core may have multiple encounters, while those in the halo may have none.   In addition (as seen above), some stars may get into loosely bound multiples \citep[cf.][]{Moeckel:Clarke:2011} and potentially have numerous self-encounters which can inflate the encounter rate significantly (see above).

\begin{table*}
    \renewcommand{\arraystretch}{1.2}
    \centering
    \caption{
        For each cluster in the $N=300$, $r_{\rm h} = 0.5$ pc ensemble (column 1) with $N$ stars (column 2), we show the half-mass radius $r_{\rm h}$ (column 3), crossing time $t_{\crossingtime}$ (column 4), and the total number $N_{\stars}$ (column 5) and fraction $f_{\stars}$ (column 6) of stars that have had an encounter within $1000$ au in $10$ Myr.  Half mass radii and crossing times are calculated explicitly for each cluster.  The final row shows the means and variances of each quantity over the ensemble.
    }\label{TAB:ENCOUNTER-FRACTIONS}
    \begin{tabular}{@{ }lccccccccc@{ }}
        \hline\hline
        Cluster & $N$ & $r_{\halfmass}$ & $t_{\crossingtime}$ & $N_{\stars}$ & $f_{\stars}$  \\ \midrule

        \texttt{N3SMR050-A} & $ 300$ & $0.510$ & $ 1.06$ & $ 162$ & $ 0.54$\\
        \texttt{N3SMR050-B} & $ 300$ & $0.509$ & $ 1.06$ & $ 163$ & $ 0.54$\\
        \texttt{N3SMR050-C} & $ 300$ & $0.505$ & $ 1.05$ & $ 143$ & $ 0.48$\\
        \texttt{N3SMR050-D} & $ 300$ & $0.500$ & $ 1.03$ & $ 155$ & $ 0.52$\\
        \texttt{N3SMR050-E} & $ 300$ & $0.499$ & $ 1.03$ & $ 160$ & $ 0.53$\\
        \texttt{N3SMR050-F} & $ 300$ & $0.501$ & $ 1.03$ & $ 164$ & $ 0.55$\\
        \texttt{N3SMR050-G} & $ 300$ & $0.499$ & $ 1.03$ & $ 170$ & $ 0.57$\\
        \texttt{N3SMR050-H} & $ 300$ & $0.505$ & $ 1.05$ & $ 153$ & $ 0.51$\\
        \texttt{N3SMR050-I} & $ 300$ & $0.507$ & $ 1.05$ & $ 144$ & $ 0.48$\\
        \texttt{N3SMR050-J} & $ 300$ & $0.499$ & $ 1.03$ & $ 157$ & $ 0.52$\\ \hline
        \texttt{N3SMR050-A..J} & $ 300$ & $0.503\pm0.004$ & $ 1.04\pm 0.01$ & $157\pm9$ & $ 0.52\pm 0.03$\\
        \bottomrule
    \end{tabular}
\end{table*}

Despite the large variation in encounter rates, When we measure the encounter {\em fraction} in each of our ten clusters we find that this value is statistically the same.  In Table \ref{TAB:ENCOUNTER-FRACTIONS} we show the number ($N_{\rm s}$) and fraction ($f_{\rm s}$) of stars in each of the ten statistically identical clusters from Fig. \ref{FIG:ENCOUNTER-RATES-N3SM} and Table \ref{TAB:NUMBER-DENSITY} that have had an encounter within 1000 au after 10 Myr.  This number is between 143 and 170 ($157 \pm 9$) -- statistically consistent with being the same number, and a little over half the stars in the cluster at $f_{\rm s} = 0.58 \pm 0.03$.

This tells us that in all clusters in this ensemble {\em the same fraction of stars are having very different numbers of encounters}.

We can also examine other ensembles of statistically identical clusters and we find that the encounter fraction in different clusters in the same ensemble is statistically the same.

The mean and variances of encounter fractions for each ensemble are given in Table \ref{TAB:ENCOUNTER-FRACTIONS-ALL}, but to summarise for $r_{\rm e} < 1000$ au: for $N=300$ clusters with $r_{\rm h} = 0.75$pc, the encounter fraction is $0.38 \pm 0.02$; for $N=300$ clusters with $r_{\rm h} = 1$pc, the encounter fraction is $0.28 \pm 0.03$; and for $N=600$ clusters with $r_{\rm h} = 0.5$pc, the encounter fraction is $0.60 \pm 0.02$.

We do not present the data here in detail, but the same is true for different encounter distances: ie. the encounter fraction is lower when the distance is e.g. 500au, but the encounter fraction is statistically the same within each ensemble.  (It is difficult to say anything about extremely close encounters as we are into small-$N$ statistics.)

That the encounter fraction is constant is extremely interesting, as the most useful measure of encounters is often {\em how many} stars have had at least one encounter closer than a particular distance over a particular timescale.

\subsubsection{Encounter fractions in different clusters}

As we saw above, within an ensemble of statistically identical clusters the fraction of stars that have at least one encounter within 1000 au within 10 Myr is statistically the same, but it is different between different ensembles.

\begin{figure}
    \includegraphics[angle=270,width=\columnwidth]{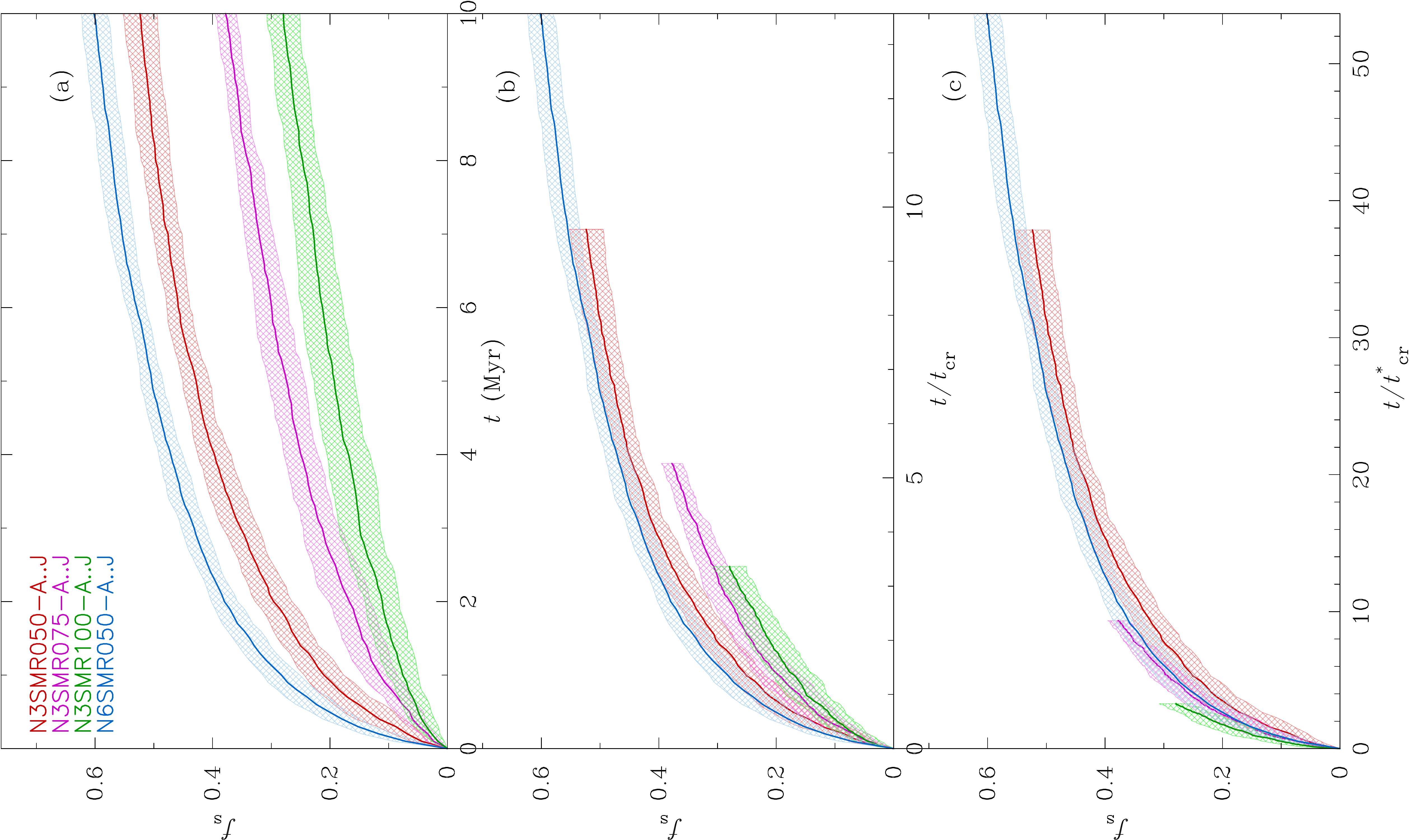}
    \caption{
        Encounter fractions for each ensemble  against the absolute time $t$ (top panel), crossing time $t_{\crossingtime}$ (middle panel), and the encounter crossing times $t_{\crossingtime}^*$ (bottom panel, see text).
        The top panel contains the legend for the IDs of each ensemble. The line is the mean value, and the shaded region shows the variance for $N=600$ clusters with $r_{\rm h} = 0.5$ pc (blue), $N=300$ clusters with $r_{\rm h} = 0.5$ pc (red), $N=300$ clusters with $r_{\rm h} = 0.75$ pc (purple), and $N=300$ clusters with $r_{\rm h} = 1$ pc (green).
    }
    \label{FIG:ENCOUNTER-FRACTIONS-3X1}
\end{figure}

\begin{table*}
    \renewcommand{\arraystretch}{1.2}
    \centering
    \caption{
        For each ensemble (column 1) with $N$ stars (column 2), we show the mean and variance of the half-mass radii $r_{\rm h}$ (column 3), crossing times $t_{\crossingtime}$ (column 4), and the total number $N_{\stars}$ (column 5) and fraction $f_{\stars}$ (column 5) of stars that have had an encounter within $1000$ au in $10$ Myr.  Half-mass radii and crossing times are calculated explicitly for each cluster.
    }\label{TAB:ENCOUNTER-FRACTIONS-ALL}
    \begin{tabular}{@{ }lccccccccc@{ }}
        \hline\hline
        Cluster & $N$ & $r_{\halfmass}$ & $t_{\crossingtime}$ & $N_{\stars}$ & $f_{\stars}$  \\ \midrule

        \texttt{N3SMR050-A..J} & $ 300$ & $0.503\pm0.004$ & $ 1.04\pm 0.01$ & $157\pm9$ & $ 0.52\pm 0.03$\\
        \texttt{N3SMR075-A..J} & $ 300$ & $0.751\pm0.002$ & $ 1.90\pm 0.01$ & $113\pm5$ & $ 0.38\pm 0.02$\\
        \texttt{N3SMR100-A..J} & $ 300$ & $1.012\pm0.021$ & $ 2.98\pm 0.09$ & $84\pm8$ & $ 0.28\pm 0.03$\\
       \texttt{N6SMR050-A..J} & $ 600$ & $0.503\pm0.004$ & $ 0.74\pm 0.01$ & $361\pm13$ & $ 0.60\pm 0.02$\\

        \bottomrule
    \end{tabular}
\end{table*}

The top panel of Fig. \ref{FIG:ENCOUNTER-FRACTIONS-3X1} shows how the encounter fraction, $f_{\stars}$ increases with (absolute) time for all of our ensembles.  The blue line and shaded region which shows the variance at the top are for the $N=600$ clusters with $r_{\rm h} = 0.5$ pc.  The red line and shaded region are for the $N=300$ clusters with $r_{\rm h} = 0.5$ pc.  The purple line and shaded region are for the $N=300$ clusters with $r_{\rm h} = 0.75$ pc.  And at the bottom the green line and shaded region are for the $N=300$ clusters with $r_{\rm h} = 1$ pc.

As can be seen, in each case the encounter fraction within ensembles evolves in the same general way  - rising rapidly and then flattening - but different ensembles seem to evolve at different rates.

That encounters occur at different rates in these different ensembles should not be a surprise as each of the clusters have a different internal dynamical timescale set by their crossing time. In the middle panel of Fig. \ref{FIG:ENCOUNTER-FRACTIONS-3X1} we show the encounter fractions by crossing time, rather than by physical time, and the differences between the different ensembles becomes less pronounced.  That the (green) $N=300$ clusters with $r_{\rm h} = 1$ pc have had the fewest encounters is clearly to a large extent because these clusters are dynamically much younger.

However, it is clearly not just dynamical age that is important as the lines are still somewhat different.  The reason for this is that the cluster size plays a role.  In all of these simulations we are counting encounters within $1000$ au which is a more significant fraction of the distances between stars in an $r_{\halfmass} = 0.5$ pc cluster than in an $r_{\halfmass} = 1$ pc cluster.  So, we would expect the encounter timescale to also be sensitive to the relative impact parameter $(r_{\halfmass}/r_{\encounter})^2$.

In the bottom panel of Fig. \ref{FIG:ENCOUNTER-FRACTIONS-3X1}, we plot encounter fraction against an `encounter crossing time', $t_{\crossingtime}^*$, defined as
\begin{equation}\label{EQN:ENCOUNTER-CROSSING-TIME}
    t_{\crossingtime}^* = t_{\crossingtime}\rbk{\frac{r_{\halfmass}/{\rm pc}}{r_{\encounter}/1000\,{\rm au}}}^2.
\end{equation}

Now in the bottom panel we appear to have found a timescale on which all clusters show extremely similar behaviour.
For encounters within $1000$ au there is a sharp rise in $f_{\stars}$ in the first $5$ $t_{\crossingtime}^*$ to a point where roughly a third of all stars have had an encounter.  It then takes another $\sim 50$ $t_{\crossingtime}^*$ for the next third of stars to have an encounter.

\begin{figure}
    \includegraphics[angle=270,width=\columnwidth]{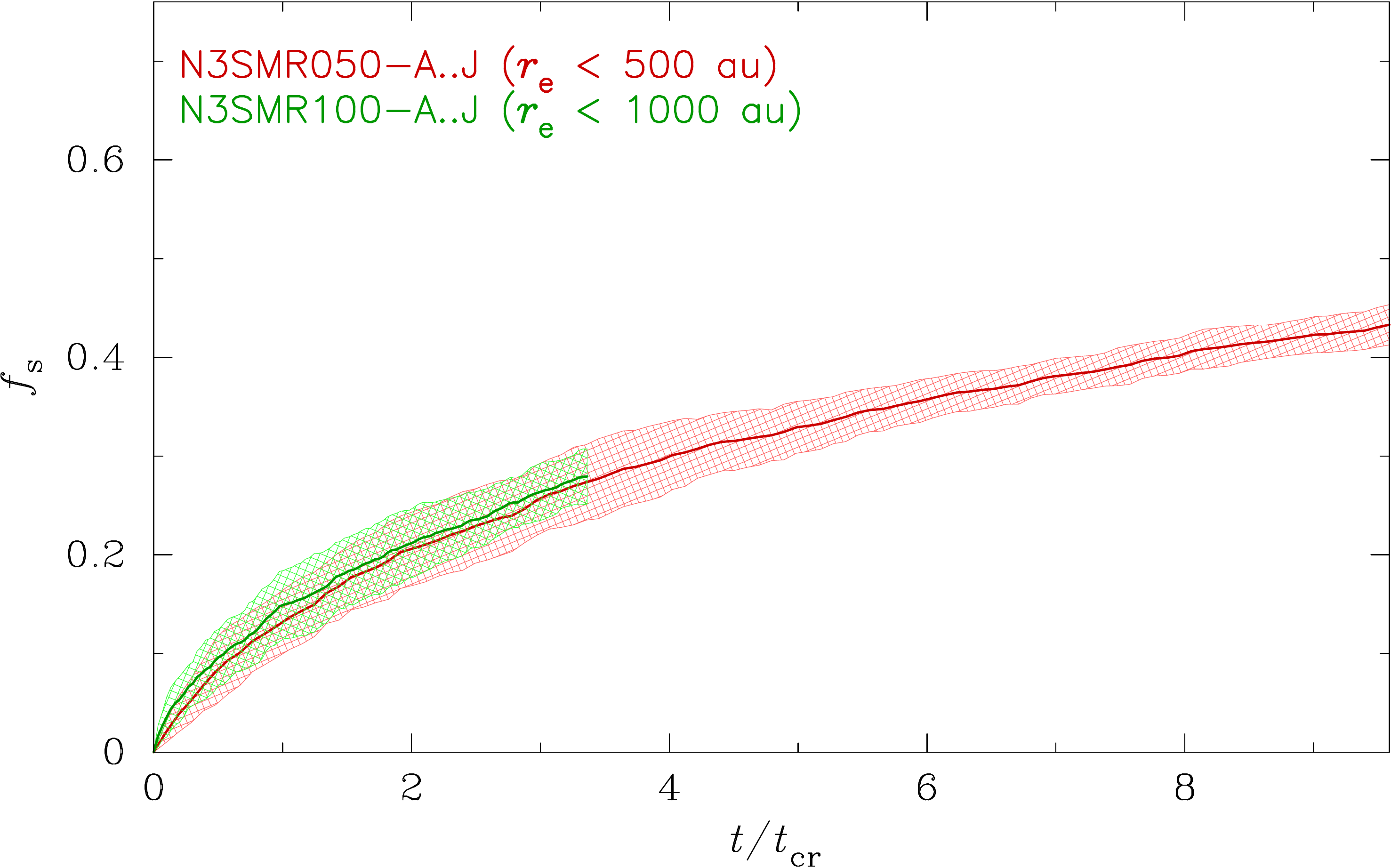}
    \caption{
        The encounter fractions against crossing time with $r_{\encounter} < 1000$ au for $N=300$, $r_{\rm h} = 1$ pc (green shaded region),  and with  $r_{\encounter} < 500$ au for $N=300$, $r_{\rm h} = 0.5$ pc (red shaded region).
    }
    \label{FIG:ENCOUNTER-FRACTIONS-re}
\end{figure}

To test the scaling with $(r_{\halfmass}/r_{\encounter})^2$, in Fig. \ref{FIG:ENCOUNTER-FRACTIONS-re} we  compare encounters within $500$ au in an $r_{\halfmass} = 0.5$ pc cluster (red) with encounters within $1000$ au in an $r_{\halfmass} = 1$ pc cluster (green).  Here $(r_{\halfmass}/r_{\encounter})^2$ is the same in both clusters (half the encounter distance, but half the half-mass radius), therefore we would expect the growth of the different encounter fractions with just the crossing time to be the same, which they are as is clear in the figure.

Note that there are various subtleties at play such as different velocity dispersions causing the effect of gravitational focusing to be different, and we are dealing with small-$N$ stochastic systems.  But overall, the overlap between evolution between different clusters within ensembles and very different ensembles is impressive when scaled by crossing time and relative encounter cross section.

\subsubsection{An empirical relationship}

The bottom panel of Fig. \ref{FIG:ENCOUNTER-FRACTIONS-3X1} provides a way of getting a rough estimate of the encounter fraction of stars, $f_{\stars}$, at a particular encounter distance, $r_{\encounter}$, after some time, $t$, in any cluster if one knows the crossing time, $t_{\crossingtime}$ and half-mass radius, $r_{\halfmass}$.  Then from equation \eqref{EQN:ENCOUNTER-CROSSING-TIME} one can calculate $t_{\crossingtime}^*$, and so $t/t_{\crossingtime}^*$.

It is worth pointing-out that the curve in the bottom panel looks like it should have a fairly simple function that would provide a fit.  However, we have struggled to find a simple ($2$ or $3$ parameter) function that fits (the problem is that the initial rise is much steeper than e.g. an exponential will fit).  Therefore we would suggest simply reading-off the value of $f_{\stars}$ from the bottom panel of Fig. \ref{FIG:ENCOUNTER-FRACTIONS-3X1} for whatever value of $t/t_{\crossingtime}^*$.

We stress that this is a rough estimate, however, as rough-and-ready as this may be, it will still almost certainly provide a {\em much} better feeling for how many stars have had an encounter than any attempt to use equation \eqref{EQN:CLUSTER-ENCOUNTER-RATE}, and then to extrapolate to an encounter fraction.

\subsubsection{An example}

We can take a roughly Orion Nebula-like cluster with $M = 1000$ M$_{\sun}$, $N = 2500$, a half-mass radius $r_{\halfmass} = 0.7$ pc, and age $3$ Myr and attempt to estimate what fraction of stars have had an encounter at $< 1000$ au.  For such a cluster, $n = 1800$ pc$^{-3}$, and assuming virial equilibrium $\sigma = 2.6$ km\,s$^{-1}$, and so $t_{\crossingtime} = 0.27$ Myr.

From equation \eqref{EQN:ENCOUNTER-CROSSING-TIME} if $r_{\encounter} = 1000$ au, then $t_{\crossingtime}^* = 0.5 t_{\crossingtime} = 0.14$ Myr.  Therefore this cluster has an `encounter age' of $t/t_{\crossingtime}^* \sim 20$.  From the bottom panel of Fig. \ref{FIG:ENCOUNTER-FRACTIONS-3X1} this would suggest around $40$ per cent of stars will have had an encounter within $1000$ au.

If we use equation \eqref{EQN:CLUSTER-ENCOUNTER-RATE}, we find $\mathcal{E} \sim 1000$ Myr$^{-1}$.  For an age of $3$ Myr this suggests $3000$ encounters among the $N = 2500$ stars.  An extremely naive extrapolation might suggest that therefore all stars had an encounter within $1000$ au.  One can be a little more sophisticated and assume that encounters are random which finds that $\sim 90$ per cent of stars will have had an encounter (as each one of the $3000$ encounters involves two stars, even if they are random most stars will have been involved in one).  Even if the value of $3000$ encounters in $3$ Myr happened, by luck, to be right, the extension of this encounter number to the number of stars involved in encounters is certainly not random.

\section{Conclusions}\label{SEC:CONCLUSIONS}

We have performed $N$-body simulations of small star clusters to investigate stellar encounters with separations $r_{\encounter} < 1000$ au.  This is the regime in which discs, planetary systems, and  multiple stellar systems can be significantly perturbed or destroyed.

We find that the encounter {\em rates} vary by up to an order of magnitude between statistically identical clusters.  However, we find that the {\em fraction} of stars that have had an encounter is statistically the same within statistically identical clusters.

The fraction of stars that have had an encounter increases rapidly at early dynamical times before flattening significantly once stars in orbits particularly susceptible to encounters have had at least one encounter.  This depends on both the dynamical timescale of the cluster ($t_{\crossingtime}$), and the relative impact parameter  $(r_{\halfmass}/r_{\encounter})^2$.

We find a consistent, and reasonably tight, relationship between the fraction of stars that have had an encounter and a modified crossing time $t_{\crossingtime}^* \propto t_{\crossingtime}(r_{\halfmass}/r_{\encounter})^2$.

The relationship we have found has a seemingly solid physical basis, but no detailed theoretical underpinning (we are working on this).  However, it provides a simple way of extracting an estimate of the encounter fraction for a particular cluster of a particular age from a figure.  While this is empirical, it almost certainly provides a {\em much} better estimate of the true encounter fraction than any attempt to apply standard theory.

\acknowledgments

SG was partly funded by STFC consolidated grant ST/V000853/1.

\bibliography{references}{}
\bibliographystyle{aasjournal}
\end{document}